\documentstyle[prl,aps]{revtex}
\begin{document}
\draft
\title{Two loops calculation in chiral perturbation theory and the
unitarization program of current algebra}
\author{J. S\'a Borges\thanks{E-mail: Borges@vax.fis.uerj.br}, J.
Soares Barbosa\thanks{E-mail: Soares@vax.fis.uerj.br} and M. D.
Tonasse\thanks{E-mail: Tonasse@vax.fis.uerj.br}}
\address{Instituto de F\'\i sica, Universidade do Estado 
do Rio de Janeiro, Rua S\~ao Francisco Xavier 524, 20550-013 -- Rio
de Janeiro, RJ\\ Brazil}
\date{\today}
\maketitle

\begin{abstract}

In this paper  we compare two loop  Chiral Perturbation Theory (ChPT)
calculation of pion-pion scattering with the unitarity second order
correction  to the current algebra soft-pion theorem.  It is shown
that both methods lead to  the same analytic structure for the  scattering
amplitude.
Current algebra; Unitarity.
\end{abstract}
\vskip 1 cm
\section{Introduction}

 In the early sixties, many results for low energy  meson physics
have been derived from the assumptions of a local chiral
SU(2)$\times$SU(2) algebra of vector and axial vector current
densities together with the partial conservation hypothesis (PCAC)
relating the derivative of the axial vector current to the pion
field.  By itself, the PCAC relation can not help one to extend the
applicability of current algebra method to the energy corresponding
to meson-meson resonances. Nevertheless, a method was invented to
obtain results for process were mesons are not ``soft particles".  We
are referring to the hard meson methods of current algebra\cite
{Wein1}.  Even ignoring the underlying theory, the chiral current
algebra implies a set of Ward identities and the method consists in
solving the system of Ward identities under  the general principles
of analyticity,  crossing and elastic unitarity.

In 1979, Weinberg suggested that it is possible to
summarize these  previous results in a phenomenological Lagrangian
which incorporates all the constraints coming from chiral symmetry of
the underlying theory\cite{Wein2}. In a set of very important and
fundamental papers, Gasser and Leutwyller have developed Chiral
Perturbation Theory (ChPT) which allows one to compute many different
Green functions involving low energy pions\cite {Leut1}. One of the
main obstacles for the applications of ChPT to high energies lies in
the issue of unitarity. Several methods have been proposed to extend
the range of energies where ChPT could be applied. These methods
include the Pad\'e expansion\cite{Dob1}, the inverse amplitude method
\cite{Dob2} and the introduction of fields describing
resonances\cite{Eck1}.

Let us remember, however  that, in order to treat pion-pion
scattering amplitude obtained by the hard meson method based on the
Ward identity technique, one of us have introduced the constraints of
elastic unitarity for partial waves\cite{Bor1}.  We will call this
approach by {\it unitarization program of current algebra} (UPCA). It
was applied to obtain first order corrections (QU1) to the soft-pion
$\pi\pi$ Weinberg amplitude\cite {Wein3} as well as to calculate
second order corrections (QU2) to the referred amplitude\cite{Bor2}.

 The ChPT calculation  of pion-pion scattering to one loop was
performed in Ref. \cite{Leut1} and, only recently, two loops
calculation appeared in the literature
\cite{Eck2}. Our aim is to compare UPCA and ChPT calculations.  In a
previous paper, we have shown that one loop ChPT   is equivalent to
QU1\cite{Bor2} and in the present paper, we will compare QU2
amplitude with ChPT Lagrangian two loops calculation performed in
Ref. \cite{Eck2}. We will conclude that, as conjectured by one of
us\cite{Bor2}, the two approaches are equivalent.  In the next
section we will recall the comparison at one loop level and in the
section three this comparison is extended to the two loops case. In
the section four we present the conclusions.

\section{First order corrected by UPCA amplitude and one loop ChPT}

Let us remember the main points of the  UPCA.  The starting point in
our derivation was an exact hard-pion expression for the correlation
function of four currents, with the quantum numbers of the pion, in
terms of three- and two-point functions.  From this expression, by
using vertex and propagator estimates, we could reobtain the
so-called soft-pion Weinberg amplitude, namely:
\begin{equation}
A^{CA}(s,t,u) = {1\over F_\pi^2}(s-m^2).
\label{eq1}
\end{equation}
The remaining of the amplitude reflects the difference between soft-
and hard-meson result. In the UPCA  one estimates the behavior of
form factors and propagators at low energies and assumes that, for
instance, for small values of its argument, the scalar  pion form
factor,\, $F_I$,\, is of the same order of magnitude as current
algebra amplitudes  near threshold.  This can be obtained by setting,
for \ $x\ \simeq\ m^2$,
\quad $F_I(x)\ \simeq \,  1\: + \: f_I^{(1)}(x)\: +\: O(\epsilon^2)$,\
for  $\: I = 0 $\ and\ $2$, \ where the superscript ``(1)" denotes
the order $\epsilon\, = \,  m^2/M^2 $,\ $M$\, being of order of
magnitude of vector meson masses.

To construct unitarized amplitude, we have observed that current
algebra gives real partial waves. The unitarization method must
provide an imaginary part to the corrected partial wave. Thus, at the
first order of the calculation, by the optical theorem, one must have

$$ {\hbox{Im}}\ t^{(1)}_{\ell I}(s) = {1\over 32\pi}\, \sigma(s)\
t^{CA}_{\ell I}(s)^2,$$

\noindent where $t^{CA}_{\ell I} $ is the soft-pion $\ell$\ partial
wave, isospin $I$,\, Weinberg amplitude obtained from Eq. (\ref{eq1})
and $$\sigma(s) = \sqrt \frac{s - 4 m^2}{s} .$$

In the program, we work with exact total amplitude expression which
follows from the Ward identities, and we use the implications of
elastic unitarity for form factors and propagators in a peculiar way.
For instance, from the relations valid for the scalar form factor and
scalar propagator, namely:

\begin{equation}
{\hbox{Im}}\ F_I(s) = \frac{1}{32\pi}\ \sigma(s)\ T^*_I(s)\  F_I(s)
\quad{\hbox{and\ \  Im}}\ \ \Delta_I(s) = \frac{1}{32\pi} \: \sigma(s)
\ \vert F_I(s)\vert^2,\ I\, = \, 0\, \mbox { and }\, 2,
\label{eq2}
\end{equation}

\noindent we obtain within the first order of the approximation:
 
$${\hbox{Im}}\  f_0^{(1)}(s) = {1\over 32\pi}\, \sigma(s)\ t^{CA}_{0
0}(s)
\quad{\hbox {and\ \  Im}}\ \ \delta_I^{(1)}(s)\, =\,
\frac{1}{32\pi}\, \sigma(s),$$

\noindent where $ t^{CA}_{0 0}$ is the current algebra isospin
zero   S-wave $\pi \pi$ amplitude, $$t^{CA}_{0 0}(s) = {1\over {
F_\pi^2}}(2 s- m^2).$$

Considering the known imaginary part of each function entering into
the amplitude, the method consists in obtaining their real parts by
the dispersion relation technique. To converge, dispersion integrals
need subtraction which are model free parameters. They can be fixed
by fitting experimental data.

In this way,  the first order corrected amplitude, QU1,  derived in
the context of UPCA\cite{Bor1}, can be  written in the following
form:

\begin{eqnarray*}
F_\pi^4 A_{_{QU1}}^{(1)} (s,t,u)& = &  {1\over 3} (2 s - m^2)
\Phi_0^{(1)}(s) - {1\over 3} (2 m^2 - s)\Phi_2^{(1)}(s)+ 
{1\over 2 }\xi_1(s-2M^2)^2+  \\ & & \left\lbrack {1\over 2} (2 m^2 -
t)\Phi_2^{(1)}(t) +(s-u)\Phi_1^{(1)}(t)- {1\over
4}\xi_2(t-2M^2)^2+(t\leftrightarrow u )\right\rbrack
\end{eqnarray*}
\noindent with:
\begin{eqnarray*}
\Phi_0^{(1)}(x)\: & = & \:  ( 2 x - m^2)(g(x) +
 \alpha_{_0}),\quad
\Phi_2^{(1)}(x)\: = \: ( 2 m^2 - x )(g(x) +\alpha_{_2}),  \\
\Phi_1^{(1)}(x) \: & = & \: {1\over 3} ( x - 4 m^2)\, g(x)
 - {1\over 3} \left ( 2\, x\, \alpha_{_1}\: -\: 4\, m^2\, g(0)\right
),
\end{eqnarray*}
\noindent where:
\begin{equation}
32\, \pi^2\, g(s)\: = \: (s- 4 m^2) \int_{4 m^2}^
\infty dx \frac
{\sigma(x)}{(x- 4 m^2)\, (x - s)}\: = \: \sigma(s) \: \ln{\frac
{\sigma(s) - 1 }{\sigma(s) + 1 }}.
\label{eq3}
\end{equation}

On the other hand, the one loop elastic pion scattering obtained from
ChPT Lagrangian \cite{Leut1} is:
\begin{eqnarray}
F_\pi^4\  A_{_{ChPT}}^{(1)}(s,t,u)\: & = & \: \frac{1}{2} ( s^2 -
m^4)
\bar J (s)+\cr
&& \frac{1}{6}\left[  \left( t ( t - u) - 2 m^2  t + 4 m^2 u  -  2 m
^4\right) \bar J ( t ) +   (t
\leftrightarrow u )\right] +\cr
&&
\left[2 (\bar \ell_1 -4/3) ( s - 2m^2)^2 + (\bar \ell_2 - 5/6) (s^2
+ (t - u)^2)+\right.\cr && \left.12 m^2 s ( \bar \ell_4 -1 ) - 3
(\bar\ell_3 + 4 \bar\ell_4 -5 ) m^4\right]/96\pi^2.
\label{eq4}
\end{eqnarray}

We can identify the function $\bar J(x)$\ with \ $2\left[\: g(x)\,
-\, g(0)\right ]$ and we have  verified that the polynomial
coefficients of these functions are the same.  We have then concluded
that the above amplitude has the same analytical structure than QU1.
Each approach have its free parameters: the model free parameters of
QU1 are $\xi 's$\ and \ $\alpha 's $\ linear combinations and the
free parameters of ChPT are $\bar\ell 's $\ linear combinations.  In
UPCA the free parameters are subtraction constants, inherent to
dispersion relation technique, and in ChPT they come from tadpole
graphs and tree graphs of order O($p^4$).

From this comparison we have shown, in a recent letter, that  one
loop ChPT amplitude can fit experimental S- and P- waves up to the
resonance region by adjusting only two parameters, namely $\bar
\ell_1$\ and  $\bar \ell_2$\cite{Bor3}.

In the next section, we will compare the second order corrected by
UPCA amplitude, QU2,  with two loops calculation\cite{Eck2}. To do
this,  we will need  one loop partial wave corresponding to the ChPT
amplitude given above. For this, one expands the combinations with
definite isospin in the s-channel into partial waves:

$ T_I (s,t) = 32 \pi\  \sum_{\ell=0}^\infty(2 \ell + 1) P_\ell (\cos
\theta)\ t_{\ell I}^{(1)}(s), \qquad I=0,1\ {\hbox{and}}\ 2.$

Using Eq. (\ref{eq4}), the resulting one loop P-wave amplitude is:

\begin{eqnarray*}
t_{1 1}^{(1)}(s) & = & \frac {1}{18\, F_\pi^4 }(s - 4 m^2)^ 2 \bar J
(s) + \left \{ \frac {m^4}{8} \Bigl( \frac{s^2}{2} -\frac{13}{16}s
m^2-m^4 \Bigr) \frac {L^2(s)}{(s-4 m^2)^2}\right.\\ &&\left. + \left(
\frac{s^3}{288} -\frac{1}{18} s^2 m^2 + \frac{1}{4} s m^4
-\frac{m^6}{8}
\right) \frac {L(s)}{\sigma (s-4 m^2)} \right.\\
&&\left.- \frac{1}{ 864\ (s-4 m^2)} \Bigl (s^3 + 37 s^2 m^2 - 149 s
m^4+ 120  m^6 \Bigr)\right.\\ &&\left. + \frac{s-4 m^2}{288}\left[  (
2 \, \bar \ell_2 - 2\, \bar \ell_1 -1)\, s +8 m^2
\right]\right\}\frac{1}{F_\pi^4\pi^2},
\end{eqnarray*}

\noindent  the isospin\ $I=0$ \ one loop\ S-wave is:

\begin{eqnarray*}
t_{0 0}^{(1)} (s)\: & = & \: \frac {1}{2 F_\pi^4} (2 s - m^2)^2 \bar
J (s)+ \frac{1}{F_\pi^4 \pi^2 } \left\{ \frac{m^4}{8 } \Bigl (
s-\frac{25}{6} m^2
\Bigr) \frac{L^2(s)}{(s-4 m^2)}\right.\\
&&\left.- \Bigl (\frac{7}{144} s^2-\frac{5}{18}s\  m^2+
\frac{25}{48} m^4
\Bigr) \frac{L(s)}{\sigma }+ \left( \frac{11}{144}  \bar \ell_1 +
\frac{7}{72} \bar \ell_2 -\frac {7}{96} \right) s^2\right.\\ &&\left.
- \left(\frac{5}{18} (\bar \ell_1 +\bar \ell_2) -\frac{1}{4}
\bar\ell_4 -\frac {481}{432} \right) s m^2 +
\frac{m^4}{18} \left( \frac{11}{2}\bar\ell_1 +7\bar \ell_2 - \frac{45}{16}
\bar\ell_3 -\frac{9}{4}\bar\ell_4 + \frac {355}{16} \right) \right\}
\end{eqnarray*}

\noindent and the isospin\ $I = 2$ \ one loop \ S-wave  is:

\begin{eqnarray}
t_{0 2}^{(1)}\: & = & \: \frac{1}{2F_\pi^4}(2m^2-s)^2\bar J(s)+
\frac{1}{F_\pi^4\pi^2} \left\{
-\frac{m^4}{16}\left(s+\frac{m^2}{3}
\right)\frac{L^2(s)}{(s-4m^2)}\right.\cr
&&\left. - \left(\frac{11}{288}s^2-\frac{1}{9}sm^2+
\frac{m^4}{48}\right)\frac{L(s)}{\sigma}\right.\cr
&&\left. -\left(\frac{1}{72}\bar \ell_1+\frac{1}{18}\bar \ell_2 +
\frac{5}{192}  \right)   s^2 -\left(\frac{1}{36}\bar
\ell_1+\frac{7}{36}\bar \ell_2
+\frac{1}{8}\bar\ell_4 -
\frac{527}{864} \right) s m^2\right.\cr
&&\left. +\left(\frac{1}{18}\bar \ell_1 +\frac{2}{9}\bar \ell_2
-\frac{1}{16}\bar\ell_3 +\frac{1}{4}\bar\ell_4 -
\frac{1}{9}\right) m^4 \right\}.
\label{eq5}
\end{eqnarray}

In these expressions we have included the contributions from  $\bar
J(4 m^2)$\ which is lacking in the expressions of partial waves given
in the Sec. 2 of the Ref. \cite{Bor3} and we have corrected an
overall sign.

\section{Second order corrected by UPCA amplitude and two loops ChPT}

We have shown that the first order correction to the soft pion
amplitude (QU1) is equivalent to one loop ChPT scattering amplitude
and, in addition, we have given the tools for constructing the next
order unitarity corrections (QU2)\cite{Bor2}.  The formula (3.10) of
Ref. \cite{Bor2} can be written as:
\begin{eqnarray*}   
F_\pi^6\ A_{_{QU2}}^{(2)} (s,t,u) & = & {1\over 3} (2 s - m^2)
\Phi_0^{(2)}(s),\\ &&-{1\over 3} (2 m^2 - s)\Phi_2^{(2)}(s)
\left\lbrack {1\over 2} (2 m^2 - t)\Phi_2^{(2)}(t)
+(s-u)\Phi_1^{(2)}(t) +(t\leftrightarrow u )\right\rbrack.
\end{eqnarray*}

\noindent We can relate the above expression with the formula (3)
obtained as a consequence of the Goldstone  nature of the
pion\cite{Ster1}, namely: $$F_\pi^4\: W_I(s)\: = \: \frac{1}{32\pi}\:
t^{CA}_I(s)\:
\Phi_I^{(2)}(s), \qquad {\hbox { for }}\: I=\: 0,\: 2 \quad {\hbox{
and }}\quad F_\pi^4\: W_1(s)\: = \: \frac{1}{48\pi}
\: \Phi_1^{(2)}(s).$$
The functions\, $W_I(s)$ are analytic except for a cut singularity at
\mbox{ $s\geq 4 m^2$ }. Their discontinuities are directly related
with the discontinuities of the functions\, $\Phi_I(s)$.

We would like to emphasize that, the general structure of the UPCA
solution comes from  the Ward identity method and  the hard meson
technique implies that the amplitude is written in terms of form
factors and propagators.  We stress that the UPCA is based on the
{\it implications of elastic unitarity relations for form factors and
propagators, and not for partial waves themselves}.

The consequences, of using  Eq. (\ref{eq2}), for instance for scalar
form factors and propagators, to second order of the approximation,
are: $$ {\hbox{Im}}\, f_I^{(2)}(s)
\: = \:
\frac{1}{32\pi} \, \sigma(s) \: \left({\hbox{Re}} \,
t_I^{(1)}(s) + t^{CA} (s) \, {\hbox{Re}}\, f_I^{(1)}(s)\right), \quad
{\hbox{Im}}\:
\delta^{(2)}_I(s)\: = \: \frac{1}{32\pi}
 \: 2 \: {\hbox{Re}}\, f_I^{(1)}(s) $$
\noindent with $\: I = 0 \:$ and \, 2. The vector form factor and
vector propagator are obtained in a similar way.  The functions\,
$\Phi_I^{(2)}(s)$, constructed from form factors and propagators are
then discontinued on the right hand cut as follows:

$$\mbox{Im} \: \Phi_I^{(2)}(x)\: = \: \frac{1}{32\pi} \, \sigma(x) \:
2 \:
\mbox{Re} \: t_I^{(1)}(x),\quad \mbox{ for }\: I\, =\,  0,\, 1 \: \mbox{ and }
 \: 2$$

\noindent and $\mbox{ Re } t^{(1)}$\ stands for the real parts of the
functions in Eq. (\ref{eq5}).  Using dispersion relation technique we
obtain:
\begin{equation}
\Phi_I^{(2)}(s)\: =\: p_I(s)\, Z(s) + q_I(s)\, G(s) + r_I(s)\, g(s) 
+ P_I(s) \: {\hbox{ for }}\, I\, =\, 0, 1 \:{\hbox{ and }}\ 2.
\label{eq6}
\end{equation}

\noindent The polynomials $p_I(s)$,\ $q_I(s)$, \ $r_I(s)$,\ and \ $P_I(s)$,
for each value of total isospin $I$,  are given in the Appendix. The
function
\mbox {$ g(s)$ } is given in (3)and
\begin{eqnarray*}
 32\, \pi^2\, G(s) \:& = &
\: (s - 4 m^2)\int_{4 m^2}^\infty\, dx\, 
\frac{\sigma(x)\, {\hbox{Re}}\, g(x)}{(x - 4 m^2)\, (x - s)}\: = \: 
\frac{1}{64\pi^2}\, \sigma^2(s)\, L^2(s)\, +\, \frac{\pi^2}{3}
\, \sigma^2(s),\\
(32\, \pi^2)^2\, Z(s)\:& = &\: (s - 4 m^2)\int_{4 m^2}^\infty\, dx\,
\frac{\sigma(x)}{(x - 4 m^2) ( x - s)}\, (L(x) + i\, \pi)^2
 \: = \: \frac{1}{3}\, L(s)\, (L^2(s) +\pi^2)\\
\end{eqnarray*} 
\noindent where
$$ L(s) = \ln{\frac { \sigma(s) - 1 }{ \sigma(s) + 1}}.  $$

The amplitude \, $A_{_{QU2}}^{(2)}$\, is then written in terms of
powers of \, $L(s)$\, and\,  $L(t)$\, and contains the free
parameters \, $\bar\ell_1$,
\, $\bar\ell_2$\, and the values of the subtraction constants ($\Phi_I$\,
 and its derivatives at \, $s = 4 m^2$).

On the other hand, ChPT amplitude calculated at two loops level
\cite{Eck1} with \, $m_\pi = 1 $\, is: $$F_\pi^4\,
A_{_{ChPT}}^{(2)}(s,t,u) \: = \: F^{(2)}(s) + G^{(2)}(s,t) +
G^{(2)}(s,u) ,$$
\noindent  with: 
\begin{mathletters}
\begin{eqnarray}
&F^{(2)}(s) & = \bar J(s)\left \{\frac{1}{16\pi^2} \left (
\frac{503}{108} s^3 - \frac{929}{54} s^2 +\frac{887}{27} s
-\frac{140}{9}\right) + b_1 ( 4 s - 3 ) + b_2 (s^2 + 4 s - 4)\right.
\cr && \left. +   \frac{1}{3} b_3 ( 8 s^3 - 21 s^2 + 48 s - 32) +
\frac{1}{3} b_4 ( 16 s^3 - 71 s^2 + 112 s - 48 ) \right \}\cr &&  +
\frac{1}{18} K_1(s)  \left [ 20 s^3 - 19 s^2 + 210 s -135
-\frac{9}{16}\, \pi^2 ( s - 4 ) \right ] \cr &&   + \frac{1}{32}
K_2(s) \left ( s \pi^2 -24 \right ) +\frac{1}{9} K_3(s) \left ( 3 s^2
- 17 s + 9 \right ),\\ &G^{(2)}(s,t) & = \bar J(t) \left \{
\frac{1}{16 \pi^2}
\left [ \frac{412}{27} - \frac{ s}{54} ( t^2 + 5 t + 159 ) - t \left(
\frac{267}{216} t^2 - \frac{727}{108} t + \frac{1571}{108} \right) \right ]
\right.  \cr
&&    + b_1 ( 2 - t ) +\frac{1}{3} b_2 ( t - 4 ) ( t^2 + s - 5 ) -
\frac{1}{6} b_3 ( t - 4 )^2 ( 3 t + 2 s - 8 ) \cr
&& \left. + \frac{1}{6} b_4 ( 2 s ( 3 t - 4 ) ( t - 4 ) - 32 t + 40
t^2 - 11 t^3) \right \}  \cr &&   + \frac{1}{36} K_1(t) \left [ 174 +
8 s - 10 t^3 + 72 t^2 - 185 t - \frac{1}{16}\, \pi^2 ( t - 4 ) ( 3 s
- 8 ) \right ]  \cr &&    + \frac{1}{9} K_2(t) \left [ 1 + 4 s +
\frac{1}{64} \, \pi^2 t ( 3 s - 8 ) \right ]  \cr &&    + \frac{1}{9}
K_3(t) \left ( 1 + 3 s t - s + 3 t^2 - 9 t \right ) + \frac{5}{3}
K_4(t) \left (4 - s - t \right ).
\end{eqnarray}
\label{eq7}
\end{mathletters}
In this expression $\bar J (s)\: =\: 2 \left ( g(s) - g(0)\right ) $,
and:
\begin{eqnarray*}
 K_1  & = & \frac{L^2(s)}{(16\pi^2)^2},\ (16\pi^2)^2\, K_2  =
\sigma^2\, L^2(s) - 4 ,\  (16\pi^2)^2 K_3   = \frac {1}{s \sigma}\,
L^3(s) + \frac{\pi^2}{s\, \sigma} L(s) -
\frac{\pi^2}{2},\\   K_4 & = & 
\frac{1}{s\ \sigma^2}\left ( \frac{1}{2}\, K_1 + \frac{1}{3} K_3 +
\frac{1}{16\pi^2} \, \bar J + s\,  \frac{ \pi^2 - 6 }{192\pi^2}\right ).
\end{eqnarray*}

Our strategy to compare \, $A_{_{QU2}}^{(2)}$\, with
$A_{_{ChPT}}^{(2)}$\, was to expand them in terms of \ $L(s)$\ and \
$L(t)$\ and then to confront its coefficients. {\it We have checked
that they are the same}. With respect to the polynomials we also
realize that the structure are the same, but clearly the coefficients
are differents because they have different origins. These polynomials
include the model free parameters to be used in order to fit the
available experimental data\cite{Exp1}.

\section{Conclusions}
Our aim is to compare Chiral Perturbation Theory (ChPT)  calculations
with the Unitarization Program of Current Algebra (UPCA).  In
previous works we have compared one loop ChPT with the first order
corrected by UPCA results for   pion-pion
\cite{Bor2} and for kaon-pion scattering\cite{Bor4} and we
have concluded that they  lead to {\it the same analytical structure
for the amplitudes}.

The two loops calculation of pion-pion scattering only recently
appeared in the literature\cite{Eck2}. However, the tools for
constructing second order corrections for the soft-pion current
algebra result has been presented more than ten years ago\cite{Bor2}.
In the present paper we compare these results and we show that, as it
was conjectured\cite{Bor2}, they have {\it the same analytical
structure}.  We  will shortly present the tools for constructing next
order UPCA correction to kaon-pion.  ChPT to two loops for \,
$K\pi$\, are not yet available, but   we expected that it will be
equivalent to the  UPCA result too.

In fact, the equivalence between the two approaches were expected.
In the hard-meson method one starts from the chiral symmetric Ward
identity {\it exact} result for the correlation function of four
currents  carrying pion quantum numbers. On the other hand, ChPT
describes the low energy dynamics of fields realizing non-linearly
chiral symmetry.  Our unitarization program is based in the
principles of analyticity, crossing and elastic unitarity which in
turn are inherent to a field theory  such  as  ChPT.

In the framework of the Generalized Chiral Perturbation Theory, it
was shown how to implement elastic unitarity strating from the one
loop partial waves. That procedure leads to equivalent O($p^6$) ChPT
amplitude \cite{Ster1}. However, the main difference from  our
unitarization procedure is that it uses the consequences of elastic
unitarity for the form factors and propagators {\it rather than for
the amplitudes themselves}.

 Despite the fact that ChPT is a well stablished low energy effective
theory for meson processes, we have shown here that the UPCA is a
suitable alternative.  One obstacle related to these two approaches
is how to fix the free parameters which, in principle, are related to
the parameters of the fundamental theory. At present, since this
relationship is still lacking, the two approaches are both in same
footing.  In ChPT context we have shown that the one loop parameters
\ $\bar\ell_1$\ and \ $\bar\ell_2$\ can be fixed by fitting S- and P-wave 
\, $\pi\pi$\,  phase-shifts\cite{Bor3}. The final expression of
pion-pion ChPT amplitude up to two loops diagrams has six parameters.
However, the D-wave amplitude from \, $A^{(2)}$ \, pick up an
imaginary part for \ $s\,
\geq\, 4\, m^2$ and we claim that the global fit of S-, P- and D-wave
phase shifts will allow one to fix the new free
parameters\cite{Bor5}.

\appendix
\section*{}

The polynomials multiplying the functions $g(s)$,\  $G(s)$\ and
$Z(s)$ in Eq. (\ref{eq6}) of second order correct by UPCA amplitudes
are:

\begin{eqnarray*}
\pi^2 p_0(s) & = & \left (  \frac{11}{72}\bar \ell_1 +
\frac{7}{36} \bar
\ell_2 + \frac{17}{48} \right ) s^2 - \left ( \frac{5}{9} \bar \ell_1
+\frac{5}{9} \bar \ell_2-\frac{1}{2}\bar\ell_4 - \frac{95}{72} \right
) s +\\ && \frac{11}{18} \bar \ell_1 + \frac{7}{9} \bar\ell_2
-\frac{45}{8}\bar\ell_3 - \frac{9}{2}\bar\ell_4  + \frac{373}{144},\\
\pi^2 p_1(s) & = & \frac{1}{432}\frac{1}{s-4}\left \{\left (6 \bar
\ell_1 - 6 \bar
\ell_2 - 2\right ) s^3 - \left ( 48 \bar \ell_1
- 48 \bar \ell_2 - 61 \right ) s^2 + \right.\\ &&\left. \left ( 96
\bar \ell_1 - 96 \bar\ell_2 - 197 \right ) s + 120 \right \}, \\
\pi^2 p_2(s) & = & \left (  \frac{1}{36}\bar \ell_1 + \frac{1}{9} \bar
\ell_2 + \frac{17}{96} \right ) s^2 - \left ( \frac{1}{18} \bar \ell_1
+\frac{7}{18} \bar \ell_2 +\frac{1}{4}\bar\ell_4+ \frac{61}{144}
\right ) s + \\ && \frac{1}{9} \bar \ell_1 + \frac{4}{9} \bar\ell_2
-\frac{1}{8}\bar\ell_3+\frac{1}{2}\bar\ell_4+ \frac{5}{18},\\ q_0(s)
& = & -\frac{2}{9 (s-4)}\left ( 50 s^3 -260 s^2 + 303 s - 36 \right
),\quad q_1(s)  = -\frac{8}{9(s-4)^2}\left ( 6 s^2 - 55 s + 64
\right ), \\ q_2(s) & = & -\frac{4}{9 (s-4)}\left ( 10 s^3 - 52 s^2 +
93 s - 72  \right ), \quad r_0(s) = \frac{4}{3 (s-4)} \left ( 6 s -
25 \right ),\\ r_1(s) & = & \frac{4}{3(s-4)^2} \left ( 3 s^2 - 13 s -
6  \right ),\quad r_2(s) = -\frac{4}{3 (s-4)} \left ( 3 s + 1 \right
).
\end{eqnarray*}
The polynomial part of QU2,   $P_I$, in Eq. (\ref{eq6}) are written
as:

\begin{eqnarray*}
P_I(s)\: & = & \: A_I s^2+B_I s+ C_I, \quad {\hbox{ for}}\: I\, = \,
0\:
\mbox{and}\: 2 \\
(s - 4) \: P_1\: & = &  A_1 s^3 + B_1 s^2 + C_1 s + D_1
\end{eqnarray*}
\noindent with:

\begin{eqnarray*}
\pi^4 A_0 & = & \frac{19}{2304}\bar\ell_1+ \frac{13}{1152}\bar\ell_2
+\frac{15}{1024}\bar\ell_3+\frac{11}{768}\bar\ell_4 - \frac{21}{4096}
+\frac{25\pi^2}{576} -\frac{3083\pi^4}{34560} + \frac{\pi^4}{2}
\Phi_0''(4),\\
\pi^4 A_1 & = &
-\frac{1}{1152}\bar\ell_1+\frac{1}{1152}\bar\ell_2-\frac{13}{55296}-
\frac{\pi^2}{452}-\frac{13
\pi^4}{30240}+\frac{\pi^4}{2}\Phi_1''(4),\\
\pi^4 A_2 & = &
\frac{1}{576}\bar\ell_1+\frac{7}{1152}\bar\ell_2+
\frac{1}{3072}\bar\ell_3-\frac{1}{384}\bar\ell_4+\frac{\pi^2}{576}-
\frac{131\pi^4}{8640}+\frac{\pi^4}{2}\Phi_2''(4), \\
\pi^4B_0 & =
&-\frac{61}{1152}\bar\ell_
1-\frac{37}{576}\bar\ell_2-\frac{105}{512}\bar\ell_3-\frac{59}{384}\bar\ell_4-
\frac{17\pi^2}{48}+\frac{4331\pi^4}{8640}+\pi^4\Phi'_0(4)-
4\pi^4\Phi''_0(4),\\
\pi^4 B_1& =
&\frac{1}{96}\bar\ell_1-\frac{1}{96}\bar\ell_2-\frac{29}{4608}+\frac{35\pi^2}
{288}+\frac{331\pi^4}{30240}+\pi^4\Phi''_1(4)-6\pi^4\Phi_1'''(4),\\
\pi^4 B_2& =
&-\frac{5}{576}\bar\ell_1-\frac{11}{288}\bar\ell_2-\frac{7}{1536}\bar\ell_3+
\frac{5}{384}\bar\ell_4-\frac{5\pi^2}{48}+\frac{409\pi^4}{4320}-\pi^4\Phi'_2(4)
-4\pi^4\Phi''_2(4),\\
\pi^4 C_0& =
&\frac{23}{288}\bar\ell_1+\frac{11}{144}\bar\ell_2+\frac{45}{128}\bar\ell_3+
\frac{37}{96}\bar\ell_4+\frac{13\pi^2}{18}-\frac{421\pi^4}{720}+\pi^4\Phi_0(4)
-4\pi^4\Phi'_0(4)+8\pi^4\Phi''_0(4),\\
\pi^4 C_1& =
&-\frac{1}{24}\bar\ell_1+\frac{1}{24}\bar\ell_2-\frac{67}{1152}-\frac{67\pi^2}
{72}+\frac{53\pi^4}{2520}+\pi^4\Phi_1'(4)-8\pi^2\Phi_1''(4)+24\pi^4
\Phi_1'''(4),\\
\pi^4 C_2& =
&\frac{1}{144}\bar\ell_1+\frac{1}{18}\bar\ell_2+\frac{5}{384}\bar\ell_3-
\frac{1}{96}\bar\ell_4+\frac{7\pi^2}{18}-\frac{163\pi^4}{720}+\pi^4\Phi_2(4)-
4\Phi'_2(4)+8\pi^4\Phi''_2(4),\\
\pi^4 D_1& =
&\frac{1}{18}\bar\ell_1-\frac{1}{18}\bar\ell_2-\frac{157}{1728}+\frac{52\pi^2}
{27}-\frac{253\pi^4}{840}.
\end{eqnarray*}

In order to compare the two approaches we have not included the
dependence on \ $1/F_\pi^8$\ in the parameters \ $b_i$. In this way
the quantities\ $b_i$\ that we have used in Eqs. (\ref {eq7}) stand
for:

\begin{eqnarray*}
b_1 & = &  8 \bar \ell_1 +2 \bar \ell_3 - 2 \bar \ell_4 +
\frac{1}{48\pi^2}\left ( 7 \ln{\frac{m_\pi}{\mu}}+
\frac{13}{6}\right ), \\
b_2 &  = & - 8 \bar \ell_1 +2 \bar \ell_4 -
\frac{1}{12\pi^2}\left (  \ln{\frac{m_\pi}{\mu}}+
\frac{1}{6}\right ), \\
b_3 & = & 2 \bar \ell_1 +\frac{1}{2} \bar \ell_2
-\frac{1}{16\pi^2}\left ( \ln{\frac{m_\pi}{\mu}}+
\frac{7}{12}\right ),\\
b_4 &  = &  \frac{1}{2} \bar \ell_2 -\frac{1}{48\pi^2}\left (
\ln{\frac{m_\pi}{\mu}}+
\frac{5}{12}\right ),\\ 
b_5 &  = & \frac{1}{16\pi^2} \left [
-\frac{31}{6}\bar\ell_1-\frac{145}{36}\bar\ell_2+\frac{7}{864}+
\frac{1}{16\pi^2}\left (\frac{625}{144} \ln{\frac{m_\pi}{\mu}}-
\frac{66029}{20736}\right)\right]-\frac{21}{16}k_1-\frac{107}{96}k_2
+r^r_5, \\ b_6 &  = & \frac{1}{16\pi^2} \left [
-\frac{7}{18}\bar\ell_1-\frac{35}{36}\bar\ell_2+\frac{1}{432}+
\frac{1}{16\pi^2}\left (\frac{257}{432} \ln{\frac{m_\pi}{\mu}}-
\frac{11375}{20736}\right)\right]-\frac{5}{48}k_1-\frac{25}{96}k_2
+r^r_6,
\end{eqnarray*}
noindent where: $$k_1 = \frac{1}{192 \pi^4}\, \ln
{\frac{m_\pi}{\mu}}\, \left ( \bar\ell_1 + \ln
{\frac{m_\pi}{\mu}}\right )\qquad k_2 = \frac{1}{96 \pi^4}\, \ln
{\frac{m_\pi}{\mu}}\, \left ( \bar\ell_2 + \ln
{\frac{m_\pi}{\mu}}\right )$$
\noindent and $\: \mu$\ is the renormalization mass scale.

\end{document}